\documentclass{article}

\usepackage{PRIMEarxiv}

\usepackage[utf8]{inputenc} 
\usepackage[T1]{fontenc}    
\usepackage{hyperref}       
\usepackage{url}            
\usepackage{booktabs}       
\usepackage{amsfonts}       
\usepackage{nicefrac}       
\usepackage{microtype}      
\usepackage{lipsum}
\usepackage{fancyhdr}       
\usepackage{graphicx}       

\usepackage[table]{xcolor}
\usepackage{cite}
\usepackage{adjustbox}
\usepackage{amsmath,amssymb,amsfonts}
\usepackage{algorithmic}
\usepackage{graphicx}
\usepackage{textcomp}
\usepackage{booktabs}
\graphicspath{{media/}}     

\NewDocumentCommand{\tsub}{m}{_{\text{#1}}} 

\title{No More Sliding Window: \\Efficient 3D Medical Image Segmentation with Differentiable Top-\textit{K} Patch Sampling
}

\author{
  Young Seok Jeon \thanks{First author} \\
  Institute of Data Science \\
  National University of Singapore \\
  Singapore\\
  \texttt{youngseokjeon74@gmail.com} \\
   \And
  Hongfei Yang \\
  Saw Swee Hock School of Public Health \\
  National University of Singapore \\
  Singapore\\
  \texttt{hfyang@nus.edu.sg} \\
   \And
  Huazhu Fu \\
  Agency for Science, Technology and Research\\
  Singapore\\
  \texttt{hzfu@ieee.org} \\
   \And   
  Yeshe Kway\\
  Centre for Clinical Magnetic Resonance Research \\  
  University of Oxford \\
  United Kingdom \\
  \texttt{yeshekway@outlook.com} \\
   \And   
  Mengling Feng \thanks{Corresponding author} \\
  Saw Swee Hock School of Public Health \\
  National University of Singapore \\
  Singapore\\
  \texttt{ephfm@nus.edu.sg} \\
  }

\begin{document}
\maketitle

\begin{abstract}
3D models surpass 2D models in CT/MRI segmentation by effectively capturing inter-slice relationships. However, the added depth dimension substantially increases memory consumption. While patch-based training alleviates memory constraints, it significantly slows down the inference speed due to the sliding window (SW) approach.
We propose No-More-Sliding-Window (NMSW), a novel end-to-end trainable framework that enhances the efficiency of generic 3D segmentation backbone during an inference step by eliminating the need for SW. NMSW employs a differentiable Top-k module to selectively sample only the most relevant patches, thereby minimizing redundant computations. When patch-level predictions are insufficient, the framework intelligently leverages coarse global predictions to refine results.
Evaluated across 3 tasks using 3 segmentation backbones, NMSW achieves competitive accuracy compared to SW inference while significantly reducing computational complexity by 91\% (88.0 to 8.00 TMACs). Moreover, it delivers a 9.1× faster inference on the H100 GPU (99.0 to 8.3 sec) and a 11.1× faster inference on the Xeon Gold CPU (2110 to 189 sec).
NMSW is model-agnostic, further boosting efficiency when integrated with any existing efficient segmentation backbones. The code is available: \href{https://github.com/Youngseok0001/open_nmsw}{https://github.com/Youngseok0001/open\_nmsw}.
\end{abstract}

\keywords{Deep Learning \and 3D Medical Image Segmentation \and Differentiable Top-\textit{K} Sampling \and Gumbel-Softmax Trick \and Efficient Infernce \and Sliding Window Inference.}

\section{Introduction}
\label{sec:introduction}
\subsection{2D vs 3D model for 3D Medical Image Segmentation}
Volumetric (3D) segmentation is a key challenge in medical imaging, with numerous applications such as tumor detection~\cite{ronneberger2015u}, volume estimation~\cite{grainger2018deep}, and controlled medical image generation~\cite{konz2024anatomically}.
Both 2D and 3D UNet-like encoder-decoder models can be applied to solve 3D segmentation tasks, though they come with distinct advantages and limitations.

For 2D models, the 3D input is destructured into a batch of 2D slices, typically along the axial direction, with predictions made independently for each slice. While 2D models are computationally efficient, they fail to capture inter-slice relationships, potentially reducing the accuracy of the final whole-volume prediction.

In contrast, 3D models analyze the complete volumetric input without structural decomposition. By distributing model weights along the axial plane, the models capture inter-slice relationships, resulting in improved accuracy compared to 2D counterparts, as shown in prior studies~\cite{avesta2023comparing,Luo_2022}. However, this structural fidelity comes at the expense of significantly higher computational complexity, limiting their practicality in resource-constrained settings.

\begin{figure}[t]
     \centering
         \includegraphics[width=0.8\columnwidth]{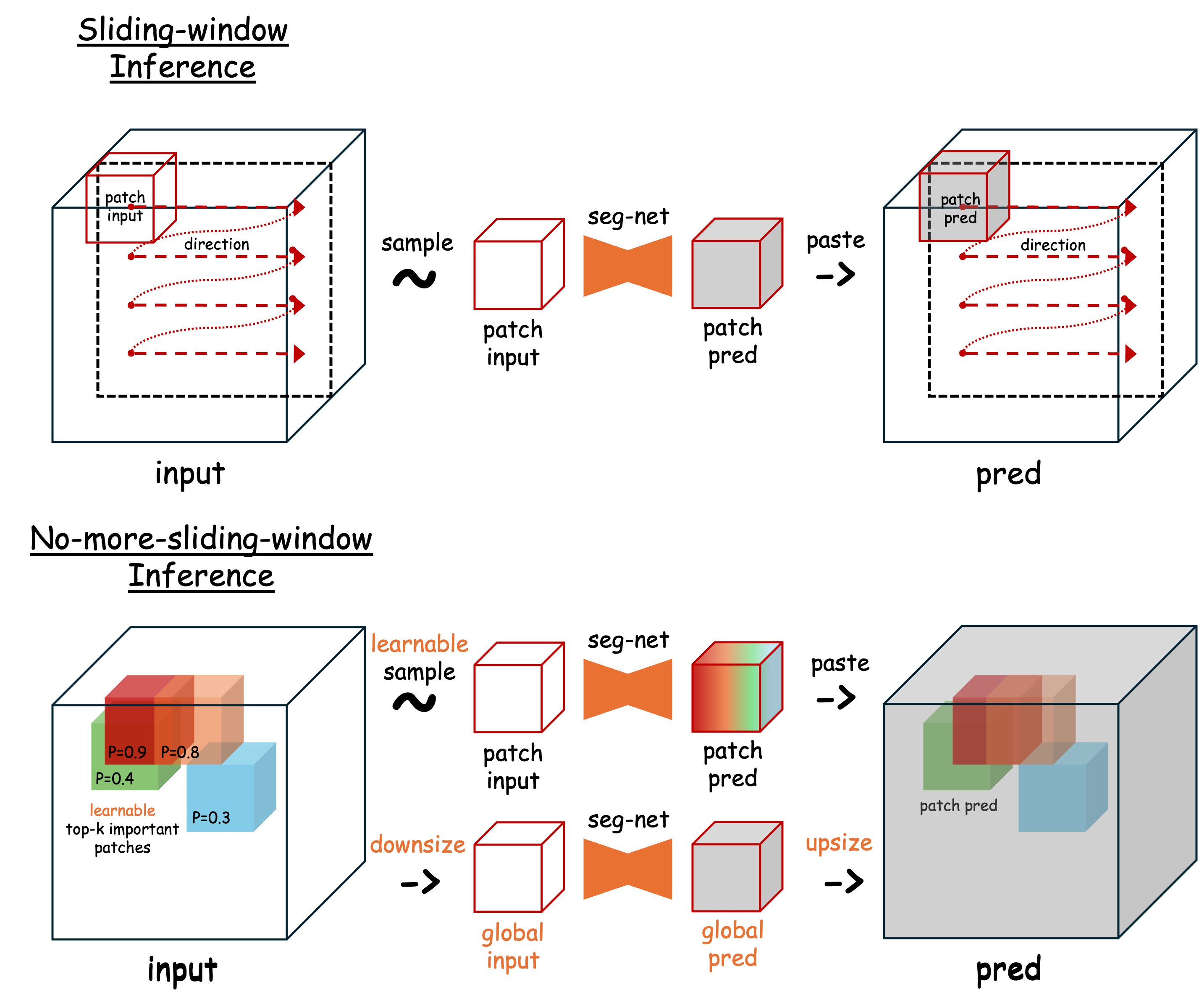}
    \caption{NMSW provides an efficient alternative to the computationally intensive Sliding Window inference method, which uniformly allocates resources across all patches. Instead, NMSW first generates a coarse whole-volume prediction and then employs a differentiable patch sampling module to automatically select a small subset of patches that require finer segmentation.}
        \label{fig:abstract}
\end{figure}

\begin{table}[t!]
\caption{Inference speed of SW across different backbones and processing units for an input of spatial size $480 \times 480 \times 480$ with 20 output classes, using a patch size of $128 \times 128 \times 128$ and an overlapping ratio of 0.5.}
\centering
\begin{adjustbox}{width=0.5\columnwidth}
\small
\begin{tabular}{lccc}
    \toprule
    \bfseries Backbone & \bfseries H100 (s) & \bfseries A100 (s)  & \bfseries Xeon Gold (s) \\
    \midrule
    UNet & 4.58 & 7.93  & 135 \\
    Swin-unetr & 24.4  & 43.4 & 2110 \\
    MedNext & 32.2 & 53.2 & 1720 \\
    \bottomrule
\end{tabular}
\end{adjustbox}
\label{table:speed1}
\end{table}

\subsection{Computational Ineffciency of 3D Segmentation Models}
Increased computational demand and memory requirements present a significant bottleneck for all 3D segmentation models, regardless of their underlying backbone—whether Conv-net~\cite{lecun1998gradient}, Self-attention~\cite{vaswani2017attention}, or State-Space models~\cite{gu2023mamba}.
Using ConvNet as an example, increasing the convolutional kernel dimension from \(k \times k\) to \(k \times k \times k\) amplifies the number of multiplications and additions required to produce each output pixel/voxel by a factor of \(k\). More critically, the memory required to store intermediate tensor outputs during backpropagation also increases, as it now includes an additional depth dimension. This challenge becomes particularly pronounced when using parallelizable matrix multiplication-based convolution algorithms, such as im2col~\cite{chellapilla2006high} or Winograd convolutions~\cite{winograd1980arithmetic}.

Transformer models are no exception to the challenges of increased memory usage and computational costs. While their architecture remains largely consistent when transitioning from 2D to 3D, the addition of tokens in the axial dimension increases the size of the self-attention matrix quadratically. For example, if a token is worth a patch of size $16 \times 16 \times 16$ (as suggested by~\cite{dosovitskiy2020image}), a CT scan of size $512 \times 512 \times 512$ yields 32,768 tokens ($32 \times 32 \times 32$), resulting in a self-attention matrix of size $32,768 \times 32,768$. 

\subsection{Sliding Window to Reduce 3D Segmentation Model's Memory}
Patch-based training, coupled with Sliding-Window (SW) inference, is currrently the predominant method to address the substantial memory requirements of 3D models. 
Rather than processing the entire volume in a single step, patch-based training randomly samples a set of patches that are significantly smaller than the actual volume.
To generate the final whole-volume prediction using the patch-trained model, SW, as shown in Figure~\ref{fig:abstract}, is employed. SW sequentially predicts patches sampled at uniform intervals with some degree of overlap (typically 25–50\%~\cite{isensee2021nnu}). The overlapping patch predictions are then aggregated into the final whole-volume prediction using an appropriate post-processing step, such as weighting each patch with a discretized Gaussian distribution to minimize edge artifacts.

\subsection{Inefficiency and Potential Prediction Bias in SW Inference}
Although SW is memory-efficient with constrained input patch size, it has significant drawbacks, including slower inference speed and increased computational demands. Because the patch-based model is not trained to rank patch importance, SW adopts a conservative approach, segmenting all patches with equal computational resources, even those that clearly do not contain objects of interest or the object is simple enough for a less complex model to handle equally well.

A typical ensemble UNet model~\cite{isensee2021nnu} takes nearly a minute to perform a full SW on a whole-volume size of $512 \times 512 \times 458$ using an NVIDIA GeForce RTX 3090 GPU~\cite{wasserthal2023totalsegmentator}. In addition, healthcare professionals often use open-source visualization tools such as 3D Slicer~\cite{fedorov20123d}, which rely on on-device processors for model inference. In such settings, inference speed drops significantly, exceeding 10 minutes (Table~\ref{table:accuracy}).

SW is not only slow during inference, but could potentially lead to poorer performance due to small patch size.
One unique aspect of medical image segmentation, compared to generic segmentation tasks on natural images, is the need for global features, such as location, which greatly help in distinguishing objects that may appear similar at a local level. Patch-trained models, however, may miss these global aspects. The most common approach to address this inherent limitation in patch-based models is to train another model that segments using a low-resolution input~\cite{isensee2021nnu}. However, this adds additional complexity to a model that is already computationally expensive.

SW is not only slow during inference, but could potentially lead to poorer performance due to a small patch size.
One unique aspect of medical image segmentation, compared to generic segmentation tasks on natural images, is the need for global features, such as location, which greatly help in distinguishing objects that may appear similar at a local level. Patch-trained models, however, may miss these global aspects. The most common approach to address this inherent limitation in patch-based models is to train another model that segments using a low-resolution input~\cite{isensee2021nnu}. However, this adds additional complexity to a model that is already computationally expensive.

\subsection{How do Radiologists Segments}
SW inference is merely a heuristic approach designed to mitigate the memory constraints of 3D segmentation models, but it does not reflect how radiologists actually segment or localize objects. Investigating radiologists' segmentation protocols could provide valuable insights for developing a more computationally efficient inference method. Unlike SW inference, radiologists break the problem down into multiple levels, each addressing different aspects of the segmentation process:
\begin{enumerate}
    \item \textbf{Initial Assessment}: Radiologists begin by quickly scanning through slices to understand the anatomical context and identify regions of interest (ROIs). During this step, they also assess and rank the segmentation difficulty of each ROI.
    
    \item \textbf{Preliminary Segmentation}: They then segment the ROIs that are easier to delineate, often using semi-automated tools powered by machine learning or traditional labeling techniques.
    
    \item \textbf{Focused Refinement}: Finally, radiologists allocate additional time to refine the coarse segmentation from the previous step, prioritizing ROIs based on the rankings established during the initial assessment.
\end{enumerate}
This workflow allows radiologists to optimize both efficiency and accuracy, ensuring a balanced trade-off between speed and precision.

\subsection{No-More-Sliding-Window}
\sloppy We introduce, for the first time in the community, a computationally efficient full-res CT/MRI segmentation framework, called the No-More-Sliding-Window (NMSW) which replaces the costly SW inference with a differentiable patch sampling module that learns to sample only a handful of patches of higher importance. NMSW aggregates the predictions from the selected patches with a low-res global prediction to produce the final full-res whole-volume prediction.

Specifically, NMSW operates through a three-step process:
(1) \textbf{Global Prediction}: A global model processes a low-resolution whole-slide volume, generating two outputs: a coarse global segmentation prediction and a probability mass function (pmf) that indicates the likelihood of regions enhancing the final prediction score.  
(2) \textbf{Patch Selection and Prediction}: High-resolution patches are selected based on the region highlights sampled from the pmf using our proposed Differentiable-Top-$K$ sampling module. These selected patches are then processed by a local model to generate granular local predictions.  
(3) \textbf{Final Aggregation}: The coarse global prediction is combined with the top-$K$ patch predictions through our Aggregation module to produce the final prediction. 

We emphasize that NMSW is not a new segmentation backbone but a framework designed to enhance the computational efficiency of existing 3D medical image segmentation backbones. NMSW can be integrated with any 3D model that previously employed SW. Across evaluations on three multi-organ segmentation tasks using three different backbone models, NMSW consistently achieves competitive segmentation performance, and in some cases even surpasses the SW baseline, while reducing computational cost by 91\%.

\subsection{Contibution}
Our contributions are summarized as follows:

\begin{itemize}
    \item For the first time in the field, we propose an end-to-end trainable framework for whole-volume 3D segmentation, called NMSW.
    
    \item NMSW eliminates the need for expensive SW inference by implementing a differentiable top-K patch sampling technique that selects only the patches most likely to enhance segmentation accuracy.
    
    \item NMSW learns to leverages coarse global predictions when patch prediction alone is insufficient.

    \item NMSW consistently achieves competitive segmentation performance, and in some cases even surpasses the SW baseline, while reducing computational cost by 91\% and 11$\times$ faster inference speed.

\end{itemize}

\section{Related Studies}

\subsection{Practice of Sliding Window Infenrence in 3D model}
SW has become the dominant strategy for managing the high memory demands of 3D medical image segmentation in nearly all newly proposed models~\cite{isensee2021nnu, hatamizadeh2021swin, sekuboyina2021verse}. Even recent efforts to improve the efficiency of 3D segmentation models primarily focus on optimizing the backbone network~\cite{9745574,shaker2024unetrdelvingefficientaccurate,perera2024segformer3d, 9491090}, while still relying on SW during inference.

Beyond 3D segmentation, patch-based methods are widely used in applications such as whole-slide image analysis in computational pathology~\cite{lu2021data, bae2023data} and mammogram classification~\cite{lotter2017multi}.

While we present our approach as a solution to SW’s limitations in 3D segmentation, it can be readily adapted to address other mega- or gigapixel challenges with minimal modifications.

\subsection{Attemps to Reduce Computation Cost of 3D model}
As aforementioned, most of recent efforts to improve the efficiency of 3D segmentation models focus primarily on optimizing the backbone architecture, while maintaining patch-based training, with only a few studies that attempt to reduce computational costs with efficient sampling strategy but with obvious scaling issues.

\subsubsection{Optimizing Backbone Architecture}
\cite{perera2024segformer3d} addresses the computational complexity of the self-attention module by introducing a set of smaller attention modules, where each module processes a reduced number of tokens. The outputs of these resized tokens are then concatenated to restore the original token count.
\cite{alalwan2021efficient} replaces the standard multi-channel convolutions with channel-wise convolutions or $1 \times 1$ convolutions.
\cite{qin2021efficient} uses knowledge distillation to supervise the training of a small student model with a large teacher model.
Our proposed approach can be added on top of these lightweight models to make the inference cost even cheaper and faster.

\subsubsection{Optimizing Inference Stretegy}
Though not explicitly mentioned in the literature, the following works have the potential to replace SW in 3D segmentation, but with some notable scaling issues.

Mask R-CNN \cite{he2018maskrcnn} is a two-stage instance segmentation model that initially identifies ROI bounding boxes, followed by segmentation within the chosen ROIs. A low-resolution input could potentially be used in the initial region proposal stage. However, as noted in the article, the proposed ROIs frequently overlap substantially, thus a high number (300 to 1000) of crops should proceed to the segmentation stage for optimal results, potentially leading to even slower inference speed than SW.

Likewise, \cite{recasens2018learning} proposes a two-step model where the first stage estimates a deformed grid from a low-resolution input, and the second stage uses this grid for resampling~\cite{jaderberg2015spatial} the high-resolution input in an end-to-end differentiable manner. However, when applied to segmentation tasks, the image resampled from the learned grid can lead to information loss due to imprecise intensity interpolation, resulting in sub-optimal segmentation performance, when the prediction is mapped back to a uniform grid.

\cite{man2019deep} proposes using Deep Q Learning~\cite{mnih2013playing} to allow a reinforcement learning (RL) agent to iteratively update a bounding box through a finite action space (e.g., zoom, shift). The reward is based on the overlap between the proposed bounding box and the ground truth mask. The updated bounding box is then used to crop a region for segmentation. Nevertheless, due to the iterative update procedure that involves treating a 3D volume as a state, this method is computationally inefficient and struggles with scalability for multi-organ tasks.

\subsection{Differentiable Patch Sampling in Computer Vision}
While the use of a smart patch selection module instead of random patch training with SW may seem like a trivial solution that many have considered in the past, the non-differentiable nature of the sampling operation has been the key limiting factor in deep learning application.

In recent years, several methods have been proposed to make sampling operations differentiable. One of the earliest approaches to achieving differentiable sampling in deep learning was introduced by Kingma and Welling~\cite{kingma2013auto}, who applied the reparameterization trick to a normal distribution, enabling backpropagation through stochastic nodes in neural networks.

Reparameterization trick was further explored by Maddison et al.\cite{maddison2016concrete} and Jang et al.\cite{jang2016categorical}, extending differentiability to any finite discrete distribution. Their approach used the Gumbel-Max trick \cite{gumbel1954statistical,maddison2014sampling} which detaches a stochastic node from the rest of learnable determistic nodes, and replaced the discrete argmax operation with temperature annealed softmax. Besides the reparameterization trick introduced, other approaches enabling learning through stochastic sampling include optimal transport \cite{xie2020differentiable}, perturbed optimization \cite{berthet2020learning}, and Monte Carlo approximation \cite{paisley2012variationalbayesianinferencestochastic}.

Few studies have applied differentiable sampling techniques for importance-based patch sampling~\cite{cordonnier2021differentiable, katharopoulos2019processing}, primarily targeting high-resolution 2D classification tasks. In contrast, our proposed approach is specifically designed for 3D medical image segmentation, which presents additional challenges due to the high spatial resolution, small batch training, and the need to integrate coarse global predictions in the regions where predicted patches do not cover.

\begin{figure}[t]
    \centering
    \includegraphics[width=\columnwidth]{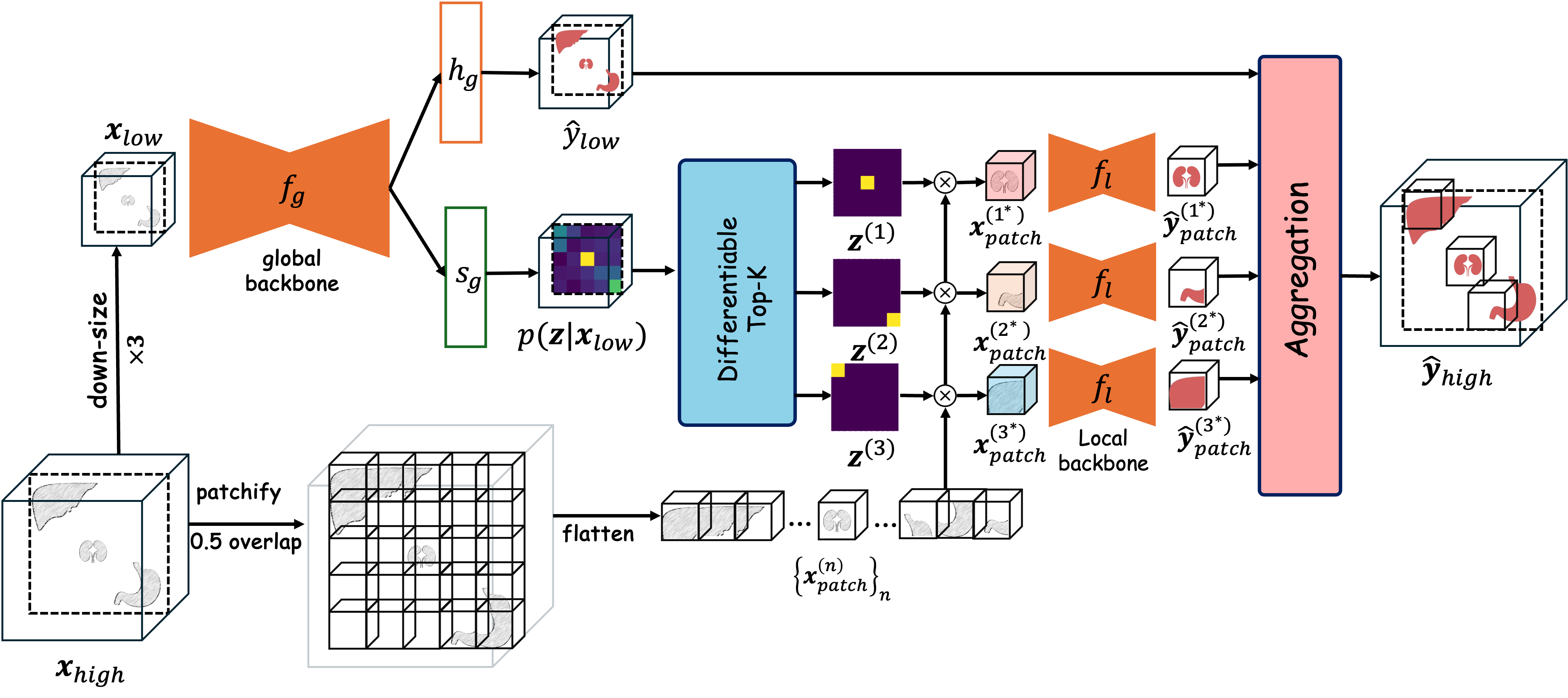}
    \caption{NMSW takes the full resolution scan $\mathbf{x}\tsub{high}$ as an input. Global backbone $f_g$ maps a downsized scan $\mathbf{x}\tsub{low}$ to a coarse global prediction $\hat{\mathbf{y}}\tsub{low}$ and a discrete probability distribution $\text{p}(\mathbf{z}|\mathbf{x}\tsub{low})$ that measures the importance of regions in the scan. $K$ regions highlights $\{\mathbf{z}^{(k)}\}$ are sampled from $\text{p}(\mathbf{z}|\mathbf{x}\tsub{low})$ with \textit{Differentiable\_Top-K} module. Region highlights are used to sample the patches in the corresponding region $\{\mathbf{x}\tsub{patch}^{(k)} \}$. Sampled patches are mapped to patch predictions $\{\hat{\mathbf{y}}\tsub{patch}^{(k)} \}$ with local backbone $f_l$.  $\{\hat{\mathbf{y}}\tsub{patch}^{(k)} \}$ and $\hat{\mathbf{y}}\tsub{low}$ are combined in \textit{Aggregation} module to produce the final Whole-Body prediction $\hat{\mathbf{y}}\tsub{high}$.
    }
    \label{fig:main}
\end{figure}

\section{Method}
\subsection{NMSW Overview}
As shown in Fig.~\ref{fig:main}, unlike conventional patch-based training, which processes a batch of patches randomly sampled from a 3D full resolution scan, NMSW takes the entire scan, $\mathbf{x}_{\text{high}} \in \mathbb{R}^{H \times W \times D}$, as input.
In the model, $\mathbf{x}\tsub{high}$ is mapped to a downsized scan $\mathbf{x}\tsub{low} \in \mathbb{R}^{H' \times W' \times D'} $ and a list of overlapping patches sampled at a regular interval $ \mathbf{X}\tsub{patch} = [\mathbf{x}^{(1)}\tsub{patch}, \mathbf{x}^{(2)}\tsub{patch}, ..., \mathbf{x}^{(N)}\tsub{patch}] \in \mathbb{R}^{N \times H_p \times W_p \times D_p} $,
where $N$ is the total number of patches \footnote{The size of $N$ is computed as $N = N_h \cdot N_w \cdot N_d$, where $N_h$, $N_w$, and $N_d$ are the patch numbers at corresponding dimension. For instance, in dimension $D$, $N_d$ is given by $N_d = \left\lfloor \frac{D - D_p}{D_p \cdot o_d} \right\rfloor + 1$, where, $r_d$ is the downsampling ratio, and $o_d$ is the overlaping ratio.}. 

The global backbone $f_g$, a generic semantic segmentation model, takes $\mathbf{x}\tsub{low}$ and produces two outputs: 1) a coarse global prediction $\hat{\mathbf{y}}\tsub{low} \in [0,1]^{C \times H_l \times W_l \times D_l}$ and 2) a discrete pdf $\text{p}( \mathbf{z} | \mathbf{x}\tsub{low} ) \in [0,1]^{N}$ (represented as 2D in Fig~\ref{fig:main} for better visualization) which estimates the importance of individual patches in contributing to the accuracy of the final segmentation map $\hat{\mathbf{y}}\tsub{high}$.

$K$ important patch locations, $\{\mathbf{z}^{(k)}\in\{0,1\}^{N}\}_{k=1}^{K}$, are sampled from $\text{p}(\mathbf{z} | \mathbf{x}_{\text{low}})$ without replacement using our \textit{Differentiable\_Top-K} module. 
Subsequently, $K$ important patches, $\{\mathbf{x}^{(k^ *)}_{\text{patch}}\}_{k^{*}=1}^{K}$, are chosen from $[\mathbf{x}^{(1)}_{\text{patch}}, \mathbf{x}^{(2)}_{\text{patch}}, \dots, \mathbf{x}^{(N)}_{\text{patch}}]$ based on $\{\mathbf{z}^{(k)}\}_{k=1}^{K}$.

The selected patches are mapped to patch predictions, $\{\hat{\mathbf{y}}^{(k)}_{\text{patch}} \in [0,1]^{C \times H_p \times W_p \times D_p}\}_{k=1}^{K}$, using another semantic segmentation model at the patch level, denoted as $f_l$. The global and local models do not share weights and are not required to have identical model architectures. As the final step, the \textit{Aggregation} module combines $\hat{\mathbf{y}}_{\text{low}}$ with $\{\hat{\mathbf{y}}^{(k)}_{\text{patch}}\}$ to produce the final whole-volume prediction, $\hat{\mathbf{y}}_{\text{high}}$. 

All modules in NMSW are differentiable, enabling end-to-end gradient-based training to produce the whole-volume prediction. Thus NMSW eliminates the need for heuristic methods like SW to convert patch predictions into a full-volume prediction.

\begin{figure}[t]
    \centering    
    \includegraphics[width=\columnwidth]{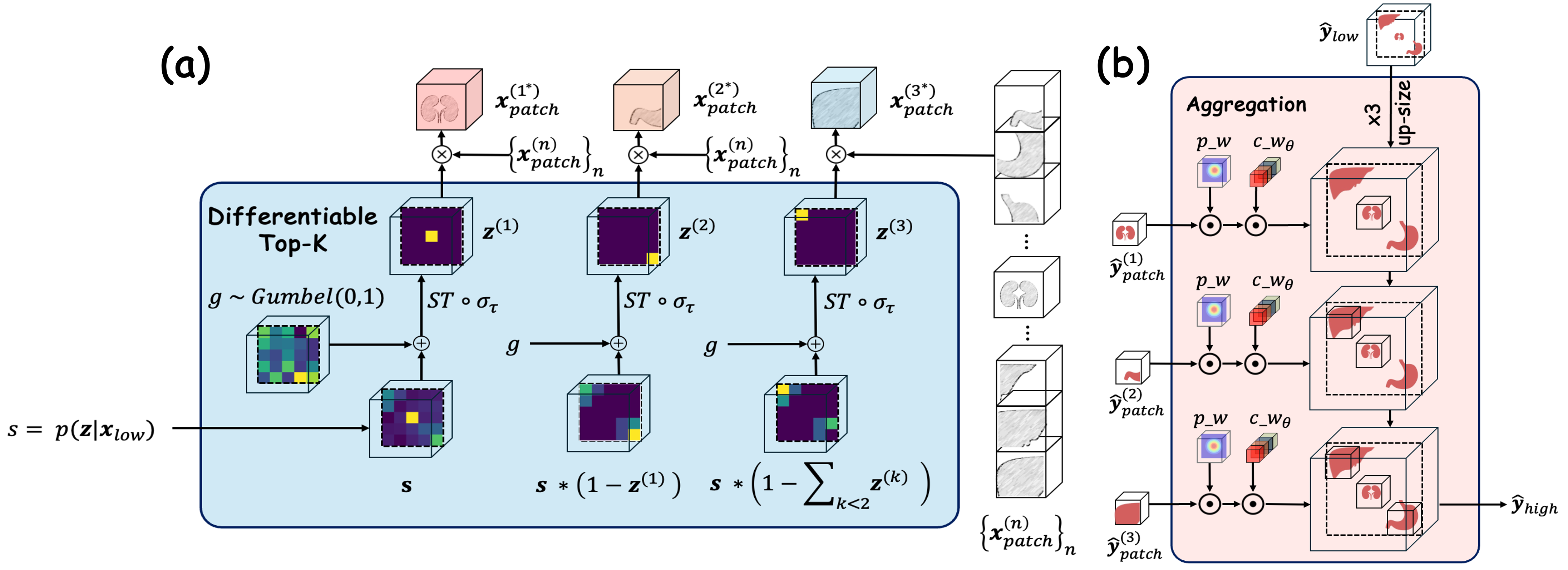}
    \caption{\textbf{Illustration of (a) Differentiable Top-K Block and (b) Aggregation Block.} 
    The \textit{Differentiable Top-K} block selects the top-K important patches $\{\mathbf{z}^{(k)}\}$ from $\text{p}(\mathbf{z}|\mathbf{x}_{\text{low}})$ using the Gumbel-Top-K trick, a differentiable extension of Gumbel-Softmax. This method iteratively modifies the distribution $\text{p}(\mathbf{z}| \mathbf{x}_{\text{low}})$ by masking the probability of previously sampled region.
    The \textit{Aggregation} block combines the global prediction $\hat{\mathbf{y}}_{\text{low}}$ and local patch predictions $\{\hat{\mathbf{y}}_{\text{patch}}^{(k)}\}$ to generate the final prediction $\hat{\mathbf{y}}_{\text{high}}$. Local predictions are weighted by $p\_w$ (overlap adjustment) and $c\_w_{\theta}$ (relative importance) before aggregation with the upscaled global prediction.
    }
    \label{fig:blocks}
\end{figure}

\subsection{Modules}
\subsubsection{Differentiable Top-K}
%
Training a stochastic module like \textit{Differentiable Top-K} is challenging due to two non-differentiable operations: (1) the random sampling operation and (2) the categorical nature of the samples.
To address the non-differentiability problem, we introduce a modified version of Reparameterazable Subset Sampling algorithm~\cite{kool2019stochastic2}, which generalizes the Gumbel-Softmax trick to a Top-K sampling scenario.


\textbf{Gumbel-Softmax~\cite{jang2016categorical,maddison2016concrete}:}
We begin by introducing Gumbel-Softmax, which proposes reparameterizable continuous relaxations of a categorical distribution.
Given a categorical distribution $\text{p}(\mathbf{z})$, where the probability of the $n$-th outcome is $p(\mathbf{z} = n) = \pi_{n}$, Gumbel-Softmax approximates the sampling operation ($z \sim p(\mathbf{z})$) as:

\begin{equation}
\label{eq:gumbel_softmax}
\mathbf{z}_{\text{soft\_hot}} = [y_1, y_2, \dots, y_N], \quad
y_n = \sigma_{\tau}(\log(\pi_{n}) + g_{n}),
\end{equation}

where taking argmax on $\mathbf{z}_{\text{soft\_hot}}$ gives $z$. $g_{i} \sim \text{Gumbel}(0, 1)$ is a sample from the Gumbel distribution with $\mu = 0$ and $\beta = 1$.\footnote{Sampling from the Gumbel distribution can be simulated by transforming uniform samples: $g = -\log(-\log(u)), \, u \sim \text{U}(0, 1)$.} 
The $\sigma_{\tau}$ is a softmax function with a temperature parameter $\tau \in [0,\infty]$, defined as:

\begin{equation}
\label{eq:softmax}
\sigma_{\tau}(x_n) = \frac{\exp((x_n/\tau)}{\sum_{m=1}^{N} \exp((x_m/\tau)}.
\end{equation}

As $\tau$ approaches 0, the Gumbel-Softmax estimator approximates the target categorical distribution $\text{p}(\mathbf{z})$. 

\textbf{Differentiable Top-K:} generalizes the Gumbel-Softmax estimator to draw Top-K samples without replacement. Unlike Gumbel-Softmax which has a fixed distribution $\text{p}(\mathbf{z})$ throughout sampling, \textit{Differentiable Top-K} (Fig~\ref{fig:blocks}(a)) masks the probabilities of previously sampled outcomes. For instance, the $k$-th random sample, $\mathbf{z}_{\text{softhot}}^{(k)}$, is defined as:

\begin{equation}
\mathbf{z}_{\text{softhot}}^{(k)} = [y_1^{(k)}, y_2^{(k)}, \ldots, y_N^{(k)}], \quad
y_i^{(k)} = \sigma_{\tau}(\log(\pi_{i}^{(k)}) + g_{i}),
\end{equation}

where 

\begin{equation}    
    \pi_{i}^{(k)} = 
    \begin{cases}
    \pi_i^{(k-1)} & i \neq \text{argmax}(\mathbf{z}_{\text{softhot}}^{(k-1)}), \\
    0 & \text{otherwise},
    \end{cases}
\end{equation}

and for the initial case, $\pi_{i}^{(1)} = \pi_{i}$.

Our application strictly requires that the sampled variable be \text{onehot} rather than \text{softhot}, as artifacts from other patches may otherwise contaminate the extracted patches. While the Straight-Through (ST)~\footnote{In pytorch ST is done with a simple gradient detach $\mathbf{z}_{\text{onehot}} := \mathbf{z}_{\text{softhot}}.\text{detach}() + \mathbf{z}_{\text{onehot}}.\text{detach}() - \mathbf{z}_{\text{softhot}}$} estimation addresses this by enabling \text{onehot} behavior in the forward pass and \text{softhot} behavior in the backward pass, it introduces huge gradient bias especially during the early training stage when the probability is not saturated. To mitigate this, we propose an adjustment to ST by scaling the \text{one\_hot} samples with their corresponding \text{soft\_hot} values: $\mathbf{z}^{(k)} = \mathbf{z}_{\text{onehot}}^{(k)} \cdot \mathbf{z}_{\text{softhot}}^{(k)}$. This adjustment accelerates convergence by reducing the gradient bias and empirically accelerates convergence. Top-K patches are extracted from the candidate patches via a simple inner product:

\begin{equation}
    \mathbf{x}_{\text{patch}}^{k^*} = \langle \mathbf{z}^{(k)}, \mathbf{X}_{\text{patch}} \rangle.
\end{equation}

Since the maximum value of $\mathbf{z}^{(k)}$ lies within the interval [0, 1], the intensity of the extracted patches is proportionally affected, as illustrated by the colored output patches in Fig~\ref{fig:blocks}(a).

Our Differentiable Top-$K$ module differs significantly from that of Cordonnier et al.\cite{cordonnier2021differentiable} in both learning objective and module design. \cite{cordonnier2021differentiable} simulates Top-$K$ sampling using perturbed maximization\cite{berthet2020learning}, which yields soft-hot samples and results in blended patches. While such blending may be tolerable for 2D classification, it deteriorates the fine-grained features crucial for 3D segmentation. In contrast, our Gumbel-Softmax-based Top-$K$ module produces clean, unblended patches better suited for segmentation task.

\subsubsection{Aggregation}
Given a set of patch predictions $\{\hat{\mathbf{y}}^{(k)}\}$ from the selected Top-K patches and the coarse global prediction $\mathbf{x}\tsub{global}$. \textit{Aggregation} block merges the two predictions into the final whole scan prediction $\hat{\mathbf{y}}\tsub{high}$ in a fully differentiable way. 

A naive approach would be to simply paste or overwrite the predicted patches onto the upscaled coarse prediction. However, this approach could lead to suboptimal predictions for two reasons: (1) Predicted patches often have overlapping areas, and merely pasting them one after the other disregards contributions from other patch predictions in these regions. (2) A simple overwrite assumes that patch predictions are always more accurate than the coarse prediction, which may not hold true, especially for objects that rely on global features.

The \textit{Aggregation} block~\ref{fig:blocks}(b) addresses the two biases present in the naive aggretation approach. Before adding the patches to the upscaled coarse prediction, each patch is multiplied by a patch weight, $p\_w \in [0,1]^{P_h \times P_w \times P_d}$, which is a discretized Gaussian distribution with $\mu=0$ and $\sigma=0.125$. This multiplication enables the model to blend the patches smoothly in the overlapping regions. \footnote{Although it is possible to make $p_w$ learnable, our experiments did not show significant improvement in performance}.
Additionally, a learnable class weight $\mathbf{c\_w}_{\theta} \in [0,1]^{C}$ is introduced to leverage global predictions for specific organs when necessary. The final prediction in patch region $\Omega(k)$ is computed as:
\begin{equation}
\hat{\mathbf{y}}_{\text{high}}(\Omega(k)) := \sigma(\mathbf{c\_w}_{\theta}) \cdot \mathbf{p\_w} \cdot \hat{\mathbf{y}}_{\text{patch}}^{(k)}  + (1- \sigma(\mathbf{c\_w}_{\theta})) \cdot \hat{\mathbf{y}}_{\text{up}}({\Omega(k)}).
\end{equation}

\subsection{Loss Function}
We apply the conventional soft-Dice combined with cross-entropy to three model predictions: $\hat{\mathbf{y}}_{\text{low}}$, $\{\hat{\mathbf{y}}_{\text{patch}}^{(k)}\}_{k=1}^{K}$, and $\hat{\mathbf{y}}_{\text{high}}$. To encourage the model to explore diverse regions during training, we introduce an entropy regularization term on the patch selection distribution $\text{p}(\mathbf{z} \mid \mathbf{x}_{\text{low}})$. The total loss function is defined as:

\begin{equation}
\label{eq:loss}
\begin{aligned}
    L_{\text{total}} = &\; L_{\text{dice}}(\mathbf{y}_{\text{low}}, \hat{\mathbf{y}}_{\text{low}}) 
    + L_{\text{dice}}(\mathbf{y}_{\text{high}}, \hat{\mathbf{y}}_{\text{high}}) \\
    &+ \frac{1}{K} \sum_{k=1}^{K} L_{\text{dice}}(\mathbf{y}_{\text{patch}}^{(k)}, \hat{\mathbf{y}}_{\text{patch}}^{(k)}) 
    + \lambda \mathcal{H}(\text{p}(\mathbf{z} \mid \mathbf{x}_{\text{low}})),
\end{aligned}
\end{equation}

where $\mathbf{y}_{\text{low}}$, $\{\mathbf{y}_{\text{patch}}^{(k)}\}_{k=1}^{K}$, and $\mathbf{y}_{\text{high}}$ represent the ground-truth labels corresponding to the respective predictions, $\mathcal{H}$ denotes the entropy function, and $\lambda \in [0,1]$ is a hyperparameter.

\begin{figure}[t]
     \centering
         \includegraphics[width=0.8\columnwidth]{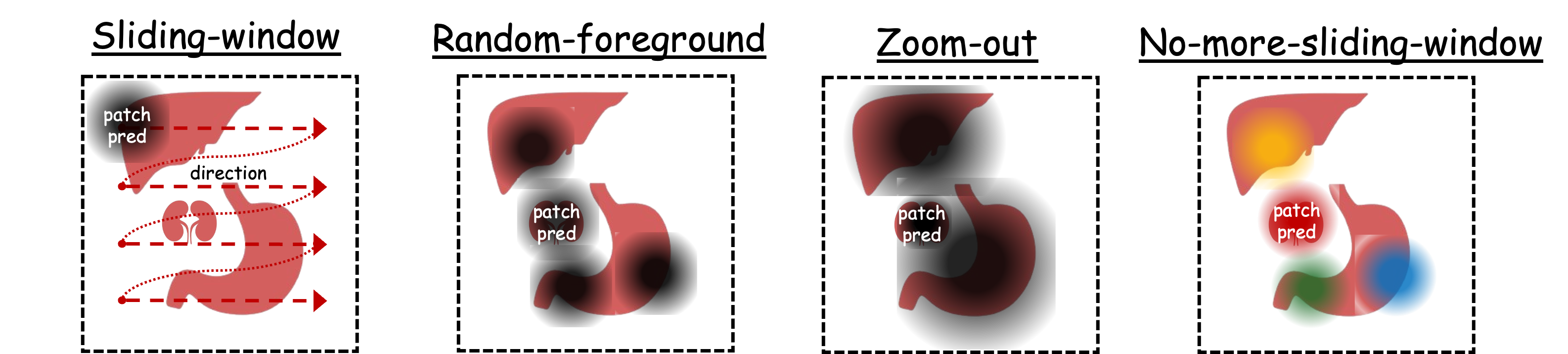}
    \caption{We evaluate NMSW in comparison to SW, Random-foreground, and Zoom-out inference techniques. The key advantage of NMSW over RF is its ability to rank patch importance, a feature absent in RF. This ranking capability is visually illustrated in the figure, where patches are highlighted in different colors to reflect their relative importance. Zoom-out is an adaptive inference strategy that generates one patch prediction per organ. If an organ exceeds the patch size, the patch is downscaled accordingly. Both Random-foreground and Zoom-out need a global model.}
        \label{fig:comprison}
\end{figure}

\section{Experiments}

\subsection{Comparative Baseline Inference Techniques}
Fig.~\ref{fig:comprison} illustrates the three baseline inference methods—SW, Random Foreground (RF), and Zoom-out—which are compared with our NMSW.
SW is the most commonly used inference technique, relying solely on a local patch-based model. It generates the final volume prediction by scanning the volume at regular intervals.
RF shares similarities with NMSW, with two key distinctions: 1) the sampling strategy and 2) the method of aggregation. Like NMSW, RF samples patches containing objects of interest. However, it does not rank their importance; all patches containing objects of interest are equally likely to be selected. Additionally, in the \textit{Aggregation block}, the class weights  $c\_w_{\theta}$ are set to a high value, assuming that local predictions typically outperform global predictions.
Zoom-out is an adaptive inference strategy that generates one patch prediction per organ. If an organ exceeds the patch size, the patch is downscaled accordingly. Both RF and Zoom-out need a global model.

\subsection{Backbones}
For the local network, we evaluate three popular backbones for medical image segmentation: UNet~\cite{ronneberger2015u}, MedNext~\cite{roy2023mednext}, and Swin-UNETR~\cite{hatamizadeh2021swin}. Meanwhile, we consistently use UNet as the global network to ensure computational efficiency.

\subsubsection{UNet}
In UNet, we set the number of intermediate channels at each down and up-sampling convolutional block to [32, 64, 128, 256, 320] with a bottlenck block with channel size of 320. Features are upsampled with simple tri-linear interpolation.

\subsubsection{MedNext}
For MedNext, we use the medium model, which provides the best balance between performance and computational cost. MedNext is an extension of ConvNext~\cite{liu2022convnet} with additional decoder blocks. Unlike UNet, MedNext uses large convolutional kernels and grouped convolution to reduce memory costs from increased network width. The MedNext-medium model has down- and up-sampling blocks with sizes [32, 64, 128, 256, 512], plus a bottleneck block with 512 channels. Refer to our code for more detail.

\subsubsection{Swin-unetr}
Swin-unetr is an extension of the Swin Transformer~\cite{liu2021swin}. Swin-unetr consists of both convolutional and Swin Transformer blocks in the encoder, which are fused during the decoding phase. The Swin Transformer block performs self-attention within a contrained regions to reduce computational cost.

\subsection{Datasets}
Each combination of the aforementioned segmentation backbones and inference techniques is tested on three multi-organ segmentation datasets: WORD~\cite{Luo_2022}, and Organ \& Vertebrae datasets from TotalSegmentator~\cite{wasserthal2023totalsegmentator}.  We did not evaluate NMSW on datasets with limited annotated classes, such as tumor or single-organ prediction tasks, because, while NMSW is applicable to these scenarios, the problem would likely simplify our top-K patch sampling module to a boring single-patch sampling module.

\subsubsection{WORD}
WORD consists of 150 CT scans, each annotated with 16 organs. Each scan contains 159–330 slices with a resolution of 512 × 512 pixels and an in-plane spacing of 0.976 × 0.976 mm. We preprocess the data to normalize the spacing to 1 × 1 × 3 mm, resulting in a shape at the 99.5\% percentile of 512 × 512 × 336.

\subsubsection{TotalSegmentator}
TotalSegmentator contains 1,204 CT scans, annotated with 104 organs. From the full set of labels, we select two subsets: Organ and Vertebrae. TotalSegmentator exhibits considerable shape variability, with dimensions ranging from 47 × 48 × 29 to 499 × 430 × 852. We normalize the spacing to 1.5 × 1.5 × 1.5 mm, resulting in a shape at the 99.5\% percentile of 373 × 333 × 644.

\subsection{Training Details}

For both baseline inference techniques and our proposed NMSW, we fixed the overlap ratio between patches at 50\% and the patch size at 128×128×128. The down-size rate of global input is set to $3\times3\times3$ for WORD and $3\times3\times4$ for TotalSegmentator.
We trained for 300 epochs, with 500 iterations per epoch. During training, for the baseline approaches, the batch size was set to 4, while for NMSW, we sampled 3 top-$K$ patches and 1 random patch per iteration, resulting in the same number of patches as baseline approaches. For the loss in Eq~\ref{eq:loss}, the weights of soft-Dice and cross-entropy is set to 0.8 and 0.2, respectively. The entropy weight $\lambda$ is set to 0.0001.
We employed the AdamW optimizer~\cite{loshchilov2017decoupled} with a learning rate of 3e-4 and weight decay of 1e-5, along with a cosine annealing learning rate scheduler that reaches the maximum learning rate at 20\% of the training iterations.
We set the regularization constant $\lambda = 1e-4$. Softmax temperature $\tau$ is annealed from 2 to 0.33 over the course of epoch. The number of Top-$K$ patches is set to 3 during training and includes 1 random patch. For baselines, 4 random patches are sampled with 2:1 fore/background ratio. patch size is set to $128\times128\times128$.
The data augmentation pipeline includes intensity scaling, affine transformations (translation, rotation, scaling), Gaussian noise and smoothing, intensity scaling, contrast adjustments, low-resolution simulations, and grid distortion.

\begin{table}[t!]
\caption{Comparison of accuracy and efficiency between the NMSW and the baseline methods in three segmentation tasks. $k$ represents the number of patches used for each inference. The speed of RF is omitted, as its network structure is nearly identical to NMSW. The memory of Zoom-out is identical to SW. The memory of RF is identical to NMSW.  The Inference speed and MACs assumes an input of size 1 x 480 x 480 x 480 with 20 output channels. Model size of Zoom-out and RF is same as NMSW. \underline{\textbf{Best}}, \textbf{2nd-best} and \underline{3rd-best} results are marked.} 
\centering 
\begin{adjustbox}{width=\textwidth}
\small
\begin{tabular}{lc>{\centering\arraybackslash}p{1.2cm} >{\centering\arraybackslash}p{1.2cm} >{\centering\arraybackslash}p{1.2cm} >{\centering\arraybackslash}p{1.2cm} >{\centering\arraybackslash}p{1.2cm} >{\centering\arraybackslash}p{1.2cm} >{\centering\arraybackslash}p{1.2cm} >
{\centering\arraybackslash}p{1.2cm} 
c c c}
    \toprule        
     & & \multicolumn{2}{c}{\bfseries Word} & \multicolumn{2}{c}{\bfseries TotalOrgan} & \multicolumn{2}{c}{\bfseries TotalVert} & \multicolumn{2}{c}{\bfseries Speed}  \\
    \cmidrule(l){3-4} \cmidrule(l){5-6} \cmidrule(l){7-8} \cmidrule(l){9-10}   \\
    \bfseries{Inference type} & & \bfseries DSC & \bfseries NSD & \bfseries DSC  & \bfseries NSD & \bfseries DSC & \bfseries NSD  & \bfseries GPU & \bfseries CPU & \bfseries MACs(T)  & \bfseries \# Param(M) & \bfseries \# Memory (GB)  \\
    \cmidrule(l){3-4} \cmidrule(l){5-6} \cmidrule(l){7-8} \cmidrule(l){9-10}  \cmidrule(l){11-11} \cmidrule(l){12-12} \cmidrule(l){13-13}  \\

    \midrule
    \rowcolor[gray]{0.9} 
    \multicolumn{13}{c}{\bfseries UNet} \\ 
    \midrule
    {SW (gold standard) } & & \textbf{0.852} & \underline{\textbf{0.906}}  & \textbf{0.868} & \textbf{0.904} & \underline{\textbf{0.884}} & \underline{0.940}  & 12.3 & 135 & 63.2 & 26.5 & 9.30 \\         
    \midrule
    {Zoom-out} & & {0.837} & {0.888} & {0.856} & {0.890} & {0.869} & {0.933} & \underline{5.19} & {144} & \textbf{3.83} & {-}\\    
    {RF (k=5)} & & {0.790} & {0.806}  & {0.789} & {0.802} & {0.684} & {0.765} & - & - & - & -  \\ 
    {RF (k=30)} & & {0.829} & {0.870}  & {0.832} & {0.856} & {0.773} & {0.851} & - & - & -  &  -\\      
    \midrule
    {NMSW (k=5)} & & {0.825} & {0.867}  & {0.841} & {0.870} & {0.832} & {0.908} & \underline{\textbf{0.147}} & \underline{\textbf{10.0}}  & \underline{\textbf{1.27}} & -\\
    {NMSW (k=30)} & & \underline{0.845} & \underline{0.894}  & \underline{0.871} & \underline{0.902} & \underline{0.880} & \textbf{0.944} & \textbf{1.07} & \textbf{23.7}  & \underline{5.85} & - \\
    {NMSW (k=full)} & & \textbf{0.852} & \textbf{0.903}  &  \underline{\textbf{0.875}} & \underline{\textbf{0.909}} & \textbf{0.883} & \underline{\textbf{0.945}} & 13.0 &  \underline{140} & 63.2 & 53.0 & 15.8 \\
    
   \midrule
    \rowcolor[gray]{0.9} 
    \multicolumn{13}{c}{\bfseries Swin-UNETR} \\ 
    \midrule
    {SW (gold standard)} & & \underline{\textbf{0.848}} & \underline{\textbf{0.897}} & \textbf{0.839} & {0.848} & \textbf{0.846} & \textbf{0.908} & 71.1 & 1050  & 69.2 & 15.5 & 12.0 \\ 
    \midrule
    {Zoom-out} & & {0.830} & {0.878} & {0.827} & \underline{0.858} & {0.832} & \underline{0.904} & \underline{8.71} & \underline{298} & \textbf{4.17} & {-}\\        
    {RF (k=5)} & & {0.781} & {0.795}  & {0.779} & {0.785} & {0.652} & {0.766} & - & - & - &  - \\ 
    {RF (k=30)} & & {0.824} & {0.858}  & {0.807} & {0.786} & {0.747} & {0.827} & - & - & -   & -\\         
    \midrule
    {NMSW (k=5)} & & {0.827} & {0.868}  & {0.832} & {0.827} & {0.789} & {0.871} & \underline{\textbf{0.832}} & \underline{\textbf{20.2}} & \underline{\textbf{1.36}} & - \\
    {NMSW (k=30)} & & \underline{0.837} & \underline{0.882}  & \underline{\textbf{0.847}} & \textbf{0.863} & \underline{0.834} & {0.903} & \textbf{6.07} &  \textbf{94.3} & \underline{6.37} & - \\
    {NMSW (k=full)} & & \textbf{0.846} & \textbf{0.895}  &  \underline{0.837} & \textbf{0.863} & \underline{\textbf{0.860}} & \underline{\textbf{0.920}} & 72.3 &  1140 & 69.2 & 42.0  & 15.8 \\

    \midrule
    \rowcolor[gray]{0.9} 
    \multicolumn{13}{c}{\bfseries MedNext} \\ 
    \midrule
    {SW (gold standard)} & & \textbf{0.860} & \textbf{0.913}  & \underline{\textbf{0.898}} & \underline{\textbf{0.928}} & \textbf{0.909} & \underline{\textbf{0.964}} & 99.2 & 2110 & 88.0  & 17.5  & 11.6\\   
    \midrule
    {Zoom-out} & & \underline{0.845} & \underline{0.898} & {0.880} & \underline{0.914} & {0.889} & {0.939} & \underline{10.1} & \underline{383} & \textbf{5.21} & {-}\\            
    {RF (k=5)} & & {0.792} & {0.807}  & {0.812} & {0.823} & {0.685} & {0.797} & - & - & - & -\\ 
    {RF (k=30)} & & {0.834} & {0.874}  & {0.864} & {0.891} & {0.743} & {0.852} & - & - & - & -\\      
    \midrule
    {NMSW (k=5)} & &{0.826} &{0.864} &{0.867} &{0.890} & {0.861} & {0.919} & \underline{\textbf{1.07}} & \underline{\textbf{35.8}} & \underline{\textbf{1.64}} & -\\
    {NMSW (k=30)} & &{0.845} &{0.892}  & \underline{0.882} &{0.910} & \underline{0.894} & \underline{0.942} & \textbf{8.37} &  \textbf{189} & \underline{8.02} & - \\

    {NMSW (k=full)} &  &\underline{\textbf{0.863}} &\underline{\textbf{0.916}} &\textbf{0.895} & \textbf{0.927} & \textbf{0.909} & \textbf{0.956} & 101 &  2152 & 88.0 & 44.1  & 15.8 \\    
    \bottomrule
\end{tabular}
\end{adjustbox}
\label{table:accuracy}
\end{table}

\section{Results}

\subsection{Trade-off between accuracy and efficiency}

Table~\ref{table:accuracy} compares NMSW against three baselines—SW, Zoom-out, and RF—in terms of segmentation performance and computational efficiency. The observed trend is consistent throughout: although NMSW doubles the model size due to the inclusion of an additional global segmentation model, it achieves significantly improved computational efficiency while maintaining competitive segmentation performance when the number of sampled patches is 30 ($k=30$). While the RF baseline is equally efficient, its patch sampling does not necessarily target regions where the global prediction is most deficient, resulting in a significantly smaller accuracy boost from the sampled patches. Zoom-out is efficient, but its accuracy remains lower and static, not improving with increased patch sampling. Notably, when NMSW samples all patches (k $\approx$ 300) like SW, it even outperforms SW. 

In terms of Multiply-Accumulate Operations (MACs), which correlate with the model's overall energy consumption, NMSW uses approximately 90\% fewer MACs compared to SW. Additionally, NMSW is, on average, about 11× faster in both CPU (Intel Xeon Gold) and GPU (H100) environments. Notably, the speed improvement becomes more pronounced as the complexity of the backbone model increases. This is because, unlike SW, NMSW incorporates additional computations from the global segmentation model, but their impact diminishes as the local backbone's complexity increases. For example, when using the UNet as the local and global backbone, the cost of global prediction (0.18 TMACs) and aggregation (0.07 TMACs) is negligible compared 30 local predictions (5.6 TMACs) in UNet backbone benchmark. The memory consumption of NMSW during inference is approximately 4-5 GB higher than that of SW, primarily due to the inclusion of the additional global network. 

It is worth mentioning that NMSW outperforms SW in the TotalSegmentatorOrgan task in terms of overall accuracy. We hypothesize that the learnable patch sampling module in NMSW not only improves computational efficiency but also enhances accuracy by focusing on regions where the model underperforms during training. Thus, our \textit{Differentiable Top-K} module can be viewed as a special kind of active learning algorithm that samples data points where the model performs poorly. In our case, the data points are patches. Further investigation of this hypothesis is left for future research.


\begin{figure}[t]
     \centering
         \includegraphics[width=0.7\columnwidth]{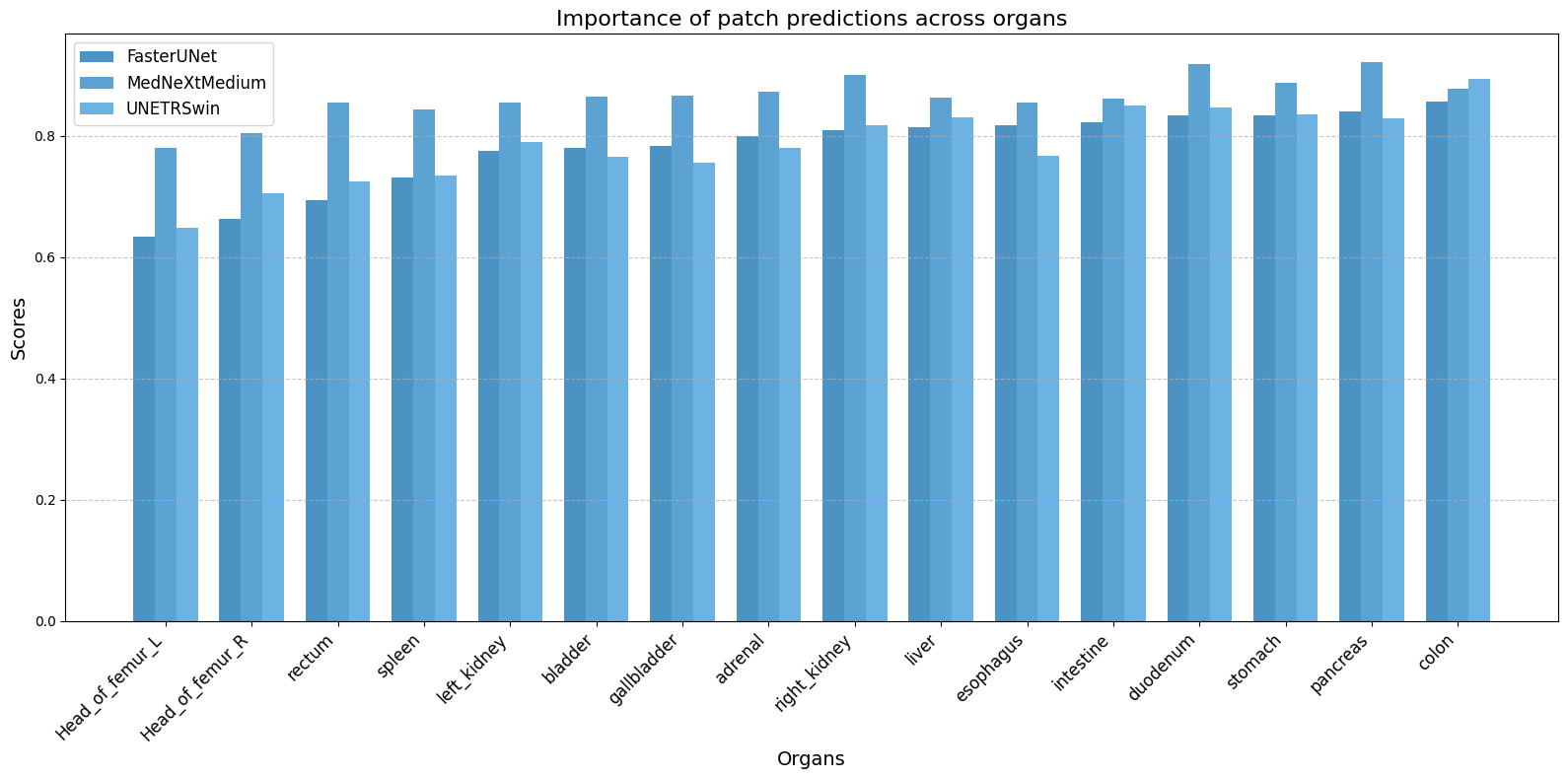}
    \caption{Class weights learned by NMSW. The model assigns higher weights (above 0.5) to patch predictions, particularly for small or complex organs (e.g., pancreas, duodenum, colon, intestine), while slightly lower weights are given to more isolated or less complex structures (e.g., femur, rectum).}
        \label{fig:class_weight}
\end{figure}

\subsection{Evaluation of the Learned Class Weight}
As shown in Fig.~\ref{fig:class_weight}, $\sigma(\mathbf{c\_w}_{\theta})$ exhibits a consistent trend across various backbones trained on the WORD task. Models generally assign higher weights (above 0.5) to patch predictions, particularly for small or complex organs (e.g., pancreas, duodenum, colon, intestine), while slightly lower weights are given to more isolated or less complex structures (e.g., femur, rectum). Interestingly, stronger backbones like MedNextMedium emphasize local predictions more heavily, effectively ignoring global predictions when the local model is powerful.

\begin{figure*}[t]
     \centering
         \includegraphics[width=0.9\textwidth]{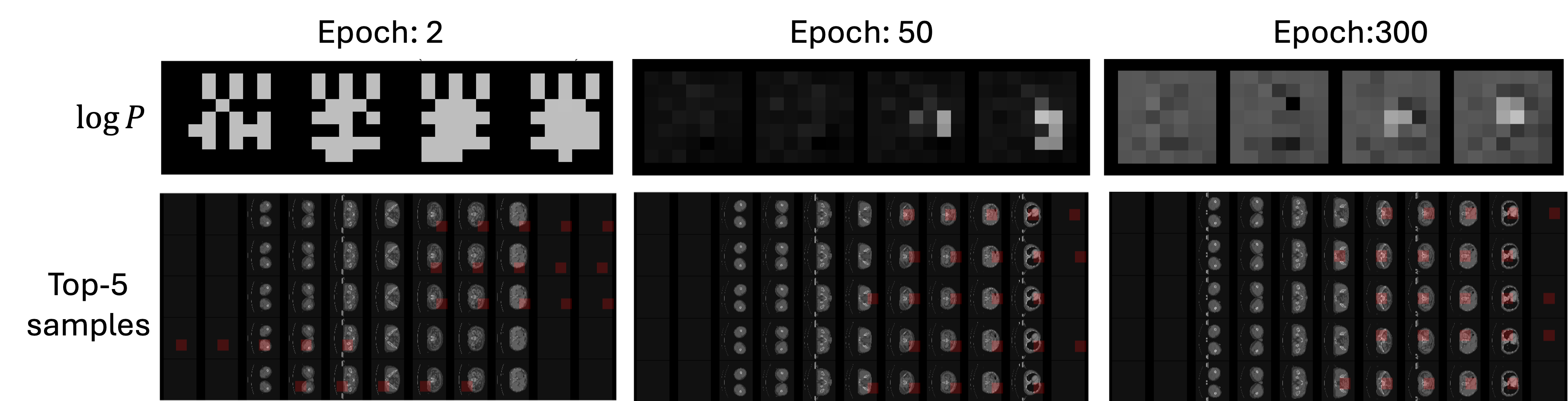}
    \caption{ the evolution of the patch sampling distribution $\text{p}(\mathbf{z}|\mathbf{x}\tsub{low})$ during training. Initially, sampling is random, ignoring foreground areas. Midway, the distribution focuses on the foreground but with significant overlap. By the end, the distribution highlights diverse foreground regions with minimal overlap.}
        \label{fig:score_evolution}
\end{figure*}

\subsection{Evaluation of sampled Top-K patches}
Fig.~\ref{fig:score_evolution} illustrates the evolution of the patch sampling distribution, $\text{p}(\mathbf{z}|\mathbf{x}\tsub{low})$, during training. Initially, the distribution is nearly random, with sampled regions (highlighted as red rectangles) scattered indiscriminately, failing to focus on foreground areas. Midway through training, the distribution begins to concentrate on foreground regions, but the sampled patches exhibit significant overlap, which is suboptimal for improving accuracy. By the end of training, the distribution not only targets foreground regions but are more well spread with less overlap. This improvement is driven by the entropy term in the loss function, which encourages exploration of diverse regions and reduces overlap among sampled patches.

\begin{figure}[t]
     \centering
         \includegraphics[width=0.7\columnwidth]{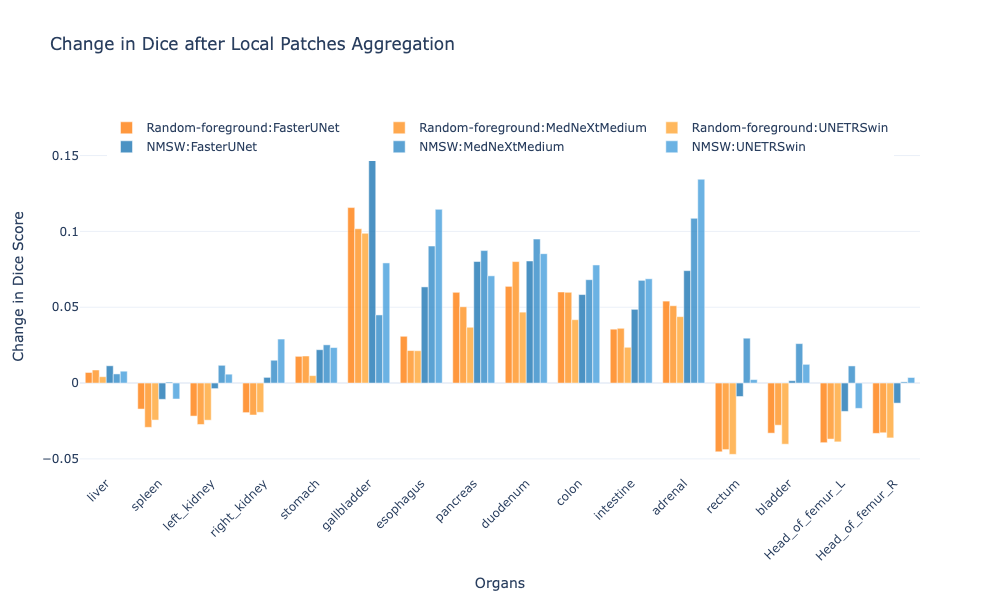}
    \caption{Dice score improvement achieved by our proposed Top-K sampling block versus the RF sampling strategy when the global prediction is supplemented with the top-5 patches.}
        \label{fig:change_in_dice}
\end{figure}

Fig.~\ref{fig:change_in_dice} compares the Dice score improvement achieved by our proposed Top-K sampling block and the RF sampling strategy when the global prediction is supplemented with the top-5 patches. While both RF and NMSW enhance performance, NMSW delivers a greater improvement across all organs. This indicates that the learned distribution is dynamic, adapting to compensate for organs where the global model underperforms, rather than focusing on a specific organ. Consequently, our sampling module is more than a simple foreground sampler with minimal overlap; it intelligently targets areas requiring enhancement.

\begin{figure}[t]
    \centering
    \resizebox{0.85\textwidth}{!}{\includegraphics{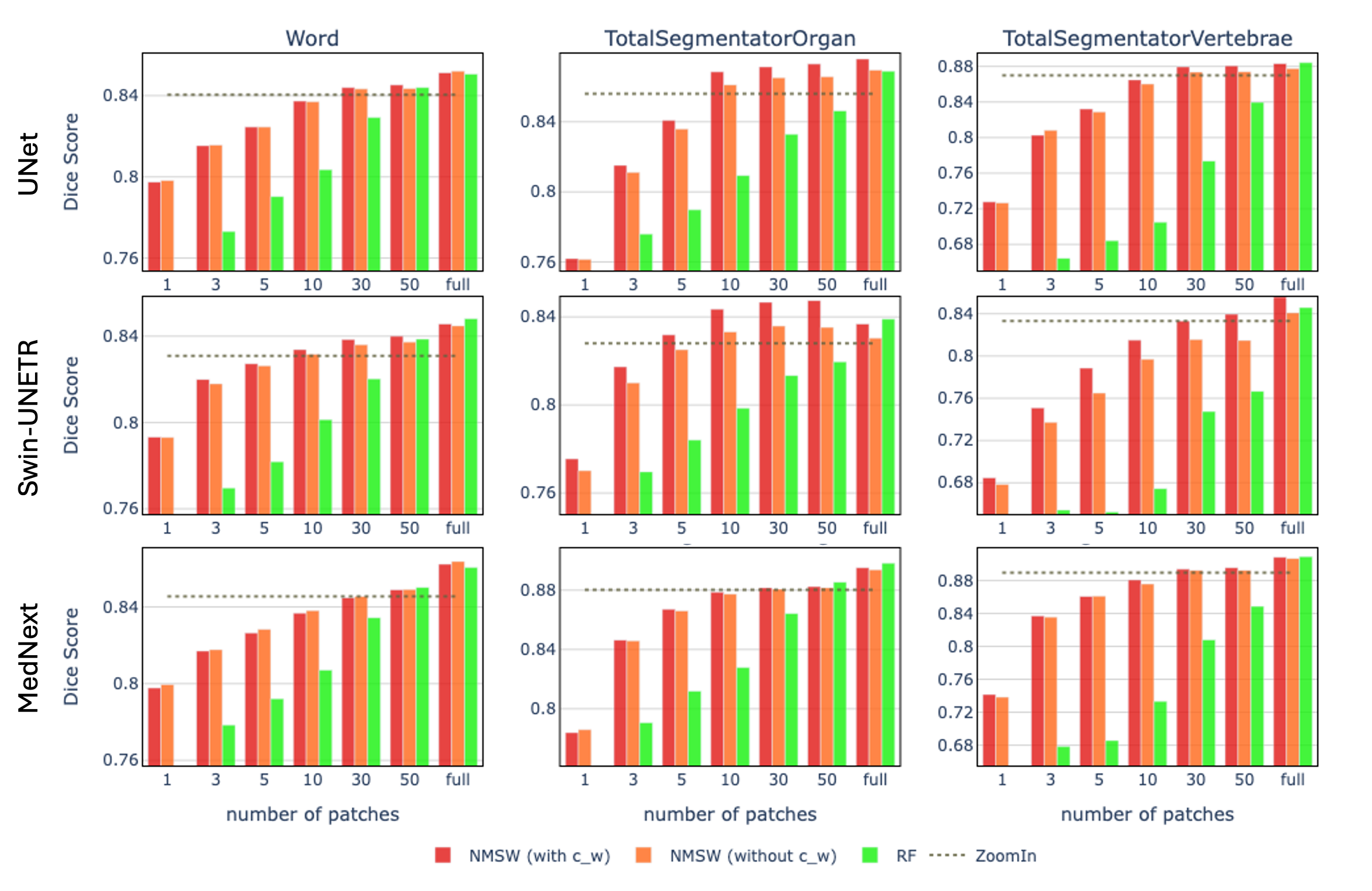}}
    \vspace{-5pt} 
    \label{fig:abil}
\end{figure}

\subsection{Ablation} 
We ablate the differentiable Top-$K$ module by replacing it with RF sampling. As shown in Figure~\ref{fig:abil}, Top-$K$ saturates around $k=30$, while RF improves linearly, saturating only when all foreground regions are sampled.  
We ablate the aggregation module by removing the class weight $c\_w_{\theta}$, forcing the model to ignore global predictions when local patches are available. Figure~\ref{fig:abil} shows that class weights generally improve the segmentation accuracy.

\section{Conclusion \& Discussion \& Future Works}
Segmentation models, like those in other fields, have become increasingly slower, larger, and more computationally expensive. While prior efforts to improve efficiency have primarily focused on simplifying backbone architectures, NMSW takes a novel approach by replacing the time-consuming sliding-window inference in 3D segmentation tasks with dynamic patch sampling. NMSW is model-agnostic and can be seamlessly integrated into existing 3D segmentation architectures with minimal computational overhead.


In evaluations across various tasks and segmentation backbones, NMSW demonstrates substantial computational savings—achieving up to 90\% lower MACs and 11$\times$ faster inference on GPUs while maintaining, and occasionally surpassing, the performance of costly SW infernce.  

While these results are promising, a few challenges remain to be addressed before the community can completely move away from SW:

\begin{itemize}
    \item \textbf{Training Speed}: Although NMSW accelerates inference, its training is slower due to the sequential dependency between global and local computations. The local network remains idle until it receives patches sampled by the global network, limiting parallelization. Future research could explore strategies to improve training efficiency by reducing this bottleneck.

    \item \textbf{Task Expansion}: We tested NMSW exclusively on instance segmentation tasks. Extending its applicability to foundation vision-language models, expending the applicability to open-vocaburary segemntation tasks.

    \item \textbf{Top-K Sampling}: The current Top-K sampling module selects patches \textit{without replacement}. However, this approach is sub-optimal if the object of interest is small enough to fit within a single patch. The remaining \( k-1 \) patches are redundant background patches. Future work could explore relaxing this restriction by enabling \textit{sampling with replacement}.
\end{itemize}

In conclusion, NMSW presents a novel approach to achieving a more compute-efficient 3D segmentation model through attention-driven patch sampling. This method stands apart from conventional approaches, which typically rely on architectural modifications to segmentation backbones that are often task-specific and non-scalable. We hope that NMSW serves as a wake-up call for the community, inspiring further exploration of dynamic sampling techniques as a pathway toward more efficient 3D medical image segmentation.

\newpage


\begin{thebibliography}{33}
\bibitem{grainger2018deep}Grainger, A., Tustison, N., Qing, K., Roy, R., Berr, S. \& Shi, W. Deep learning-based quantification of abdominal fat on magnetic resonance images. {\em PloS One}. \textbf{13}, e0204071 (2018)
\bibitem{konz2024anatomically}Konz, N., Chen, Y., Dong, H. \& Mazurowski, M. Anatomically-controllable medical image generation with segmentation-guided diffusion models. {\em International Conference On Medical Image Computing And Computer-Assisted Intervention}. pp. 88-98 (2024)
\bibitem{Luo_2022}Luo, X., Liao, W., Xiao, J., Chen, J., Song, T., Zhang, X., Li, K., Metaxas, D., Wang, G. \& Zhang, S. WORD: A large scale dataset, benchmark and clinical applicable study for abdominal organ segmentation from CT image. {\em Medical Image Analysis}. \textbf{82} pp. 102642 (2022,11), http://dx.doi.org/10.1016/j.media.2022.102642
\bibitem{avesta2023comparing}Avesta, A., Hossain, S., Lin, M., Aboian, M., Krumholz, H. \& Aneja, S. Comparing 3D, 2.5 D, and 2D approaches to brain image auto-segmentation. {\em Bioengineering}. \textbf{10}, 181 (2023)
\bibitem{lecun1998gradient}LeCun, Y., Bottou, L., Bengio, Y. \& Haffner, P. Gradient-based learning applied to document recognition. {\em Proceedings Of The IEEE}. \textbf{86}, 2278-2324 (1998)
\bibitem{vaswani2017attention}Vaswani, A. Attention is all you need. {\em Advances In Neural Information Processing Systems}. (2017)
\bibitem{gu2023mamba}Gu, A. \& Dao, T. Mamba: Linear-time sequence modeling with selective state spaces. {\em ArXiv Preprint ArXiv:2312.00752}. (2023)
\bibitem{chellapilla2006high}Chellapilla, K., Puri, S. \& Simard, P. High performance convolutional neural networks for document processing. {\em Tenth International Workshop On Frontiers In Handwriting Recognition}. (2006)
\bibitem{winograd1980arithmetic}Winograd, S. Arithmetic complexity of computations. (Siam,1980)
\bibitem{dosovitskiy2020image}Dosovitskiy, A. An image is worth 16x16 words: Transformers for image recognition at scale. {\em ArXiv Preprint ArXiv:2010.11929}. (2020)
\bibitem{shaker2024unetrdelvingefficientaccurate}Shaker, A., Maaz, M., Rasheed, H., Khan, S., Yang, M. \& Khan, F. UNETR++: Delving into Efficient and Accurate 3D Medical Image Segmentation.  (2024), https://arxiv.org/abs/2212.04497
\bibitem{isensee2021nnu}Isensee, F., Jaeger, P., Kohl, S., Petersen, J. \& Maier-Hein, K. nnU-Net: a self-configuring method for deep learning-based biomedical image segmentation. {\em Nature Methods}. \textbf{18}, 203-211 (2021)
\bibitem{wasserthal2023totalsegmentator}Wasserthal, J., Breit, H., Meyer, M., Pradella, M., Hinck, D., Sauter, A., Heye, T., Boll, D., Cyriac, J., Yang, S. \& Others TotalSegmentator: robust segmentation of 104 anatomic structures in CT images. {\em Radiology: Artificial Intelligence}. \textbf{5} (2023)
\bibitem{perera2024segformer3d}Perera, S., Navard, P. \& Yilmaz, A. SegFormer3D: an Efficient Transformer for 3D Medical Image Segmentation. {\em Proceedings Of The IEEE/CVF Conference On Computer Vision And Pattern Recognition}. pp. 4981-4988 (2024)
\bibitem{hatamizadeh2021swin}Hatamizadeh, A., Nath, V., Tang, Y., Yang, D., Roth, H. \& Xu, D. Swin unetr: Swin transformers for semantic segmentation of brain tumors in mri images. {\em International MICCAI Brainlesion Workshop}. pp. 272-284 (2021)
\bibitem{sekuboyina2021verse}Sekuboyina, A., Husseini, M., Bayat, A., Löffler, M., Liebl, H., Li, H., Tetteh, G., Kukačka, J., Payer, C., Štern, D. \& Others VerSe: a vertebrae labelling and segmentation benchmark for multi-detector CT images. {\em Medical Image Analysis}. \textbf{73} pp. 102166 (2021)
\bibitem{9745574}Deng, Y., Hou, Y., Yan, J. \& Zeng, D. ELU-Net: An Efficient and Lightweight U-Net for Medical Image Segmentation. {\em IEEE Access}. \textbf{10} pp. 35932-35941 (2022)
\bibitem{9491090}Qin, D., Bu, J., Liu, Z., Shen, X., Zhou, S., Gu, J., Wang, Z., Wu, L. \& Dai, H. Efficient Medical Image Segmentation Based on Knowledge Distillation. {\em IEEE Transactions On Medical Imaging}. \textbf{40}, 3820-3831 (2021)
\bibitem{lu2021data}Lu, M., Williamson, D., Chen, T., Chen, R., Barbieri, M. \& Mahmood, F. Data-efficient and weakly supervised computational pathology on whole-slide images. {\em Nature Biomedical Engineering}. \textbf{5}, 555-570 (2021)
\bibitem{bae2023data}Bae, K., Jeon, Y., Hwangbo, Y., Yoo, C., Han, N., Feng, M. \& Others Data-Efficient Computational Pathology Platform for Faster and Cheaper Breast Cancer Subtype Identifications: Development of a Deep Learning Model. {\em JMIR Cancer}. \textbf{9}, e45547 (2023)
\bibitem{lotter2017multi}Lotter, W., Sorensen, G. \& Cox, D. A multi-scale CNN and curriculum learning strategy for mammogram classification. {\em Deep Learning In Medical Image Analysis And Multimodal Learning For Clinical Decision Support: Third International Workshop, DLMIA 2017, And 7th International Workshop, ML-CDS 2017, Held In Conjunction With MICCAI 2017, Québec City, QC, Canada, September 14, Proceedings 3}. pp. 169-177 (2017)
\bibitem{alalwan2021efficient}Alalwan, N., Abozeid, A., ElHabshy, A. \& Alzahrani, A. Efficient 3D deep learning model for medical image semantic segmentation. {\em Alexandria Engineering Journal}. \textbf{60}, 1231-1239 (2021)
\bibitem{qin2021efficient}Qin, D., Bu, J., Liu, Z., Shen, X., Zhou, S., Gu, J., Wang, Z., Wu, L. \& Dai, H. Efficient medical image segmentation based on knowledge distillation. {\em IEEE Transactions On Medical Imaging}. \textbf{40}, 3820-3831 (2021)
\bibitem{he2018maskrcnn}He, K., Gkioxari, G., Dollár, P. \& Girshick, R. Mask R-CNN.  (2018), https://arxiv.org/abs/1703.06870
\bibitem{recasens2018learning}Recasens, A., Kellnhofer, P., Stent, S., Matusik, W. \& Torralba, A. Learning to zoom: a saliency-based sampling layer for neural networks. {\em Proceedings Of The European Conference On Computer Vision (ECCV)}. pp. 51-66 (2018)
\bibitem{jaderberg2015spatial}Jaderberg, M., Simonyan, K., Zisserman, A. \& Others Spatial transformer networks. {\em Advances In Neural Information Processing Systems}. \textbf{28} (2015)
\bibitem{man2019deep}Man, Y., Huang, Y., Feng, J., Li, X. \& Wu, F. Deep Q learning driven CT pancreas segmentation with geometry-aware U-Net. {\em IEEE Transactions On Medical Imaging}. \textbf{38}, 1971-1980 (2019)
\bibitem{mnih2013playing}Mnih, V. Playing atari with deep reinforcement learning. {\em ArXiv Preprint ArXiv:1312.5602}. (2013)
\bibitem{kingma2013auto}Kingma, D. Auto-encoding variational bayes. {\em ArXiv Preprint ArXiv:1312.6114}. (2013)
\bibitem{maddison2016concrete}Maddison, C., Mnih, A. \& Teh, Y. The concrete distribution: A continuous relaxation of discrete random variables. {\em ArXiv Preprint ArXiv:1611.00712}. (2016)
\bibitem{jang2016categorical}Jang, E., Gu, S. \& Poole, B. Categorical reparameterization with gumbel-softmax. {\em ArXiv Preprint ArXiv:1611.01144}. (2016)
\bibitem{gumbel1954statistical}Gumbel, E. Statistical theory of extreme valuse and some practical applications. {\em Nat. Bur. Standards Appl. Math. Ser. 33}. (1954)
\bibitem{maddison2014sampling}Maddison, C., Tarlow, D. \& Minka, T. A* sampling. {\em Advances In Neural Information Processing Systems}. \textbf{27} (2014)
\bibitem{xie2020differentiable}Xie, Y., Dai, H., Chen, M., Dai, B., Zhao, T., Zha, H., Wei, W. \& Pfister, T. Differentiable top-k with optimal transport. {\em Advances In Neural Information Processing Systems}. \textbf{33} pp. 20520-20531 (2020)
\bibitem{berthet2020learning}Berthet, Q., Blondel, M., Teboul, O., Cuturi, M., Vert, J. \& Bach, F. Learning with differentiable pertubed optimizers. {\em Advances In Neural Information Processing Systems}. \textbf{33} pp. 9508-9519 (2020)
\bibitem{paisley2012variationalbayesianinferencestochastic}Paisley, J., Blei, D. \& Jordan, M. Variational Bayesian Inference with Stochastic Search.  (2012), https://arxiv.org/abs/1206.6430
\bibitem{cordonnier2021differentiable}Cordonnier, J., Mahendran, A., Dosovitskiy, A., Weissenborn, D., Uszkoreit, J. \& Unterthiner, T. Differentiable patch selection for image recognition. {\em Proceedings Of The IEEE/CVF Conference On Computer Vision And Pattern Recognition}. pp. 2351-2360 (2021)
\bibitem{katharopoulos2019processing}Katharopoulos, A. \& Fleuret, F. Processing megapixel images with deep attention-sampling models. {\em International Conference On Machine Learning}. pp. 3282-3291 (2019)
\bibitem{cardoso2022monai}Cardoso, M., Li, W., Brown, R., Ma, N., Kerfoot, E., Wang, Y., Murrey, B., Myronenko, A., Zhao, C., Yang, D. \& Others Monai: An open-source framework for deep learning in healthcare. {\em ArXiv Preprint ArXiv:2211.02701}. (2022)
\bibitem{ronneberger2015u}Ronneberger, O., Fischer, P. \& Brox, T. U-net: Convolutional networks for biomedical image segmentation. {\em Medical Image Computing And Computer-assisted Intervention–MICCAI 2015: 18th International Conference, Munich, Germany, October 5-9, 2015, Proceedings, Part III 18}. pp. 234-241 (2015)
\bibitem{roy2023mednext}Roy, S., Koehler, G., Ulrich, C., Baumgartner, M., Petersen, J., Isensee, F., Jaeger, P. \& Maier-Hein, K. Mednext: transformer-driven scaling of convnets for medical image segmentation. {\em International Conference On Medical Image Computing And Computer-Assisted Intervention}. pp. 405-415 (2023)
\bibitem{liu2022convnet}Liu, Z., Mao, H., Wu, C., Feichtenhofer, C., Darrell, T. \& Xie, S. A convnet for the 2020s. {\em Proceedings Of The IEEE/CVF Conference On Computer Vision And Pattern Recognition}. pp. 11976-11986 (2022)
\bibitem{liu2021swin}Liu, Z., Lin, Y., Cao, Y., Hu, H., Wei, Y., Zhang, Z., Lin, S. \& Guo, B. Swin transformer: Hierarchical vision transformer using shifted windows. {\em Proceedings Of The IEEE/CVF International Conference On Computer Vision}. pp. 10012-10022 (2021)
\bibitem{loshchilov2017decoupled}Loshchilov, I. Decoupled weight decay regularization. {\em ArXiv Preprint ArXiv:1711.05101}. (2017)
\bibitem{kool2019stochastic2}Kool, W., Van Hoof, H. \& Welling, M. Stochastic beams and where to find them: The gumbel-top-k trick for sampling sequences without replacement. {\em International Conference On Machine Learning}. pp. 3499-3508 (2019)
\bibitem{he2016deep}He, K., Zhang, X., Ren, S. \& Sun, J. Deep residual learning for image recognition. {\em Proceedings Of The IEEE Conference On Computer Vision And Pattern Recognition}. pp. 770-778 (2016)
\bibitem{fedorov20123d}Fedorov, A., Beichel, R., Kalpathy-Cramer, J., Finet, J., Fillion-Robin, J., Pujol, S., Bauer, C., Jennings, D., Fennessy, F., Sonka, M. \& Others 3D Slicer as an image computing platform for the Quantitative Imaging Network. {\em Magnetic Resonance Imaging}. \textbf{30}, 1323-1341 (2012)
\end{thebibliography}

\end{document}